\documentclass[twocolumn,showpacs,preprintnumbers,amsmath,amssymb]{revtex4}
\usepackage{graphicx}

\begin{document}

\preprint{APS/123-QED}

\title{Scaling crossovers in activated escape of nonequilibrium systems: a resonantly driven oscillator.}

\author{Oleg Kogan}
\email{oleg@caltech.edu} \affiliation{Caltech, Pasadena CA}


\date{\today}
\begin{abstract}
The rate of metastable decay in nonequilibrium systems is expected
to display scaling behavior: i.e., the logarithm of the decay rate
should scale as a power of the distance to a bifurcation point
where the metastable state disappears. Recently such behavior was
observed and some of the earlier predicted exponents were found in
experiments on several types of systems described by a model of a
modulated oscillator. Here we establish the range where different
scaling behavior is displayed and show how the crossover between
different types of scaling occurs. The analysis is done for a
nonlinear oscillator with two coexisting stable states of forced
vibrations. Our numerical calculations, based on the the instanton
method allow the mapping of the entire parameter range of
bi-stability. We find the regions where the scaling exponents are
1 or 3/2, depending on the damping.  The exponent 3/2 is found to
extend much further from the bifurcation then were it would be
expected to hold as a result of an over-damped soft mode.  We also
uncover a new scaling behavior with exponent of $\approx$ 1.3
which extends, numerically, beyond the close vicinity of the
bifurcation point.
\end{abstract}

 \maketitle

\section{Introduction}
The problem of noise-induced escape from a metastable state has
been studied in various contexts, from nucleation to chemical
reactions to switching in nanomagnets. The escape rate often has
the Arrhenius form $W \propto \exp(-R/D)$, where $D$ is the noise
intensity (temperature, in the case of thermal noise).  Of central
interest both for theory and experiment is the activation energy
$R$. Starting with the Kramers paper \cite{Kramers} much work has
been put into calculating $R$ and the prefactor in the escape rate
for various physical systems. From this point of view, it is
particularly important to reveal generic features of the
activation energy $R$, such as scaling behavior with the system
parameters. The onset of the scaling behavior was noticed first
for systems close to thermal equilibrium \cite{Kurkijarvi1972,
Victora1989}. Scaling occurs also for systems far from thermal
equilibrium, which generally do not have detailed balance
\cite{Mark 3/2, Tretiakov2003, Knobloch1982, Maier???, 
Dykman2004}. Nonequilibrium systems may display several types of
scaling behaviors and crossovers between different scalings with
varying system parameters.

\begin{figure}[h]
\begin{center}
\includegraphics[width=6cm]{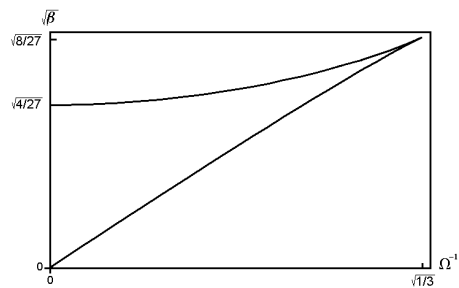}
\end{center}
\caption{\label{fig:Bistaility region}Region of bi-stability in
the parameter space of the driving, $\sqrt{\beta}$ and friction,
$\Omega^{-1}$.}
\end{figure}

\begin{figure}[h]
\begin{center}
\includegraphics[width=5cm]{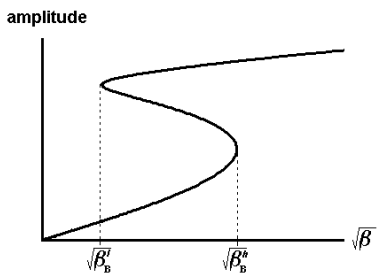} 
\end{center}
\caption{\label{fig:S-shaped figure} Amplitude versus
$\sqrt{\beta}$ for some given $\Omega^{-1} \neq 0$. The upper and
lower branches corresponds to a stable steady state, while the
middle branch corresponds to a steady state which is unstable with
respect to small perturbations. Similar curve exists for
\emph{phase} versus $\sqrt{\beta}$. A bifurcation value of
$\sqrt{\beta}$ will be denoted by $\sqrt{\beta_B}$; there are two
of these: $\sqrt{\beta_B^{h}}$ and $\sqrt{\beta_B^{l}}$.
Noise-induced transitions from low amplitude to high amplitude
branch will be called "down $\rightarrow$ up transitions", and
those from high amplitude to low amplitude branch, will be called
"up $\rightarrow$ down transitions".}
\end{figure}

A nonequilibrium system that has attracted much attention recently
is a Duffing oscillator modulated by a periodic force
\begin{equation}
\label{eq:Duffing} \ddot{q} + 2\Gamma \dot{q} + \omega_{0}^{2}q +
\gamma q^3 = A\cos{\left(\omega_F t\right)} + f(t)
\end{equation}
with high quality factor $\mathcal{Q} = \frac{\omega_0}{2\Gamma}$.
The term $f$ is white noise with property $\langle f(t)f(t')
\rangle = 2 B \delta(t-t')$. As a result of the modulation the
oscillator may have two or more coexisting states of forced
vibrations.  Noise leads to switching between these states.  In
this paper we will use the following two parameters: dimensionless
friction, $\Omega^{-1} = \Gamma/|\omega_F - \omega_0|$ and a
dimensionless driving strength, $\sqrt{\beta} = A\sqrt{3\gamma/32
\omega_F^3 |\omega_F - \omega_0|^3}$. In the space of
$\sqrt{\beta}$ and $\Omega^{-1}$, the region of bi-stability is
closed (Fig.~\ref{fig:Bistaility region}).
\\
Upon varying $\sqrt{\beta}$ at a fixed $\Omega$ one would
encounter an amplitude response curve such as that depicted in
Fig.~\ref{fig:S-shaped figure}, which exhibits a region of
bi-stability between two bifurcation points $\sqrt{\beta_B^{l}}$
and $\sqrt{\beta_B^h}$.
It is critically important to note that at any finite
$\Omega^{-1}$, bi-stability will exist only at a non-zero $\beta$.
Therefore the problem of transition rates between two stable
attractors is a \emph{non-equilibrium} problem!

Several types of scaling behavior of the switching activation
energy have been seen for such different systems as modulated
nanoelectromechanical (NEMS) \cite{Sequoia, Dynamic Range,
Buks2007_1}, and microelectromechanical (MEMS) oscillators
\cite{Chan and Stambaugh} (various sources of noise in the
nanomechanical systems have been thoroughly described in
\cite{Roukes noise}), Josephson junctions \cite{Siddiqi2005},
optically trapped atoms \cite{Jhe2005}, and superconducting
resonators \cite{Buks2007_2} - all of these systems are modelled
as Duffing oscillators. Not only is this scaling interesting on
its own, but it also underlies various applications, an important
example being quantum measurements \cite{Siddiqi2005}.

Let $\tilde{\eta} = \left|\sqrt{\beta} - \sqrt{\beta_B}\right|$
(the letter $\eta$ is reserved for denoting $\beta - \beta_B$ in
recent work by one of the authors, see \cite{Mark Quantum Scaling}
for example).  In certain cases, the switching activation energy
obeys a power law: $R \sim \tilde{\eta}^\xi$. There exist regimes
when the escape problem can be reduced to that of escape over a
1-dimensional potential barrier. Analytical approximations are
valid in two such regimes - close to bifurcation points, where
$\xi = 3/2$ and at at low $\Omega^{-1}$, where $\xi=1$ are
summarized in Section II.A. A new \emph{geometric} explanation of
these $\xi=3/2$ and $\xi=1$ exponents has recently been given
\cite{Mark and Ira}. These authors make use of a remarkable
feature of this system - that the bifurcation at
$\sqrt{\beta_B^{l}}$ is "non-local" while the bifurcation at
$\sqrt{\beta_B^{h}}$ is "local", as will be explained in Section
II.B.

In this paper we present a detailed theoretical study of the
scaling and scaling crossovers of the activation energy for a
Duffing oscillator.  Using numerical techniques, we map out the
entire bi-stability region shown in Fig.~\ref{fig:Bistaility
region}.  We identify part of this bi-stability region where the
$\xi=3/2$ and $\xi=1$ scaling laws hold, where there are
crossovers between these scaling laws, as well as regions of other
scaling behaviors not previously mentioned in the literature.
\section{Brief review of existing theory}
We will analyze the system in the canonically transformed phase
space $(Q,P)$, with $q = 2^{3/2}\omega_0\sqrt{\frac{|\omega_0 -
\omega_F|}{3\gamma}}\left(Q\cos{(\omega_F t)} + P \sin{(\omega_F
t)}\right)$ and $\dot{q} =
2^{3/2}\omega_F\omega_0\sqrt{\frac{|\omega_0 - \omega_F|
}{3\gamma}}\left(-Q\sin{(\omega_F t)} + P \cos{(\omega_F
t)}\right)$.  In the underdamped limit when $\mathcal{Q} \gg 1$,
the motion of $q$ is approximately simple-harmonic, with slow
"envelope" evolution represented by $P$ and $Q$.  The resulting
equations for $P$ and $Q$ are approximately independent of time -
the fast time-dependent corrections average to zero to lowest
order in $\mathcal{Q}^{-1}$.  In terms of slow time $T \equiv
\mathcal{Q}^{-1}t$ (subsequently, the dot denoting $d/dT$), it
follows the following evolution \cite{Envelope dynamics}:
\begin{eqnarray}
\label{eq:2D_dynamics1}\dot Q = K_Q(Q,P) + n_Q(T) \\
\label{eq:2D_dynamics2}\dot P = K_P(Q,P) + n_P(T)
\end{eqnarray}
where $K_Q(Q,P) = \frac{\partial g}{\partial P} - \Omega^{-1}Q$
and $K_P(Q,P) = -\frac{\partial g}{\partial Q} - \Omega^{-1}P$,
$g(Q,P)$ is a Hamiltonian given by $g = \frac{1}{4}(Q^2 + P^2
-1)^2-\sqrt{\beta}Q$ and $\Omega^{-1}$ is an effective damping
which tends to arrest the amplitude to zero. The $n_i$s are
stochastic terms which will be modelled as white noise: $\langle
n_i(T) n_j(T') \rangle = 2D \delta_{ij}\delta(T-T')$ where $D =
\left(\frac{3\gamma B}{16\omega_F^3\Gamma^2}\right)\Omega^{-2}$.
The terms inside $(...)$ have been defined on the first page and
are all parameters of the oscillator. In the absence of noise, an
initial condition $(Q_0, P_0)$ will evolve into one of the two
attracting fixed points (FP), or "attractors". The third FP is of
saddle type - its amplitude forms the middle branch in
Fig.~\ref{fig:S-shaped figure}. The separatrix between the basins
of attraction of each of the two attractors is a set of initial
conditions that flows into the saddle FP via its attractive
eigen-direction.

In the presence of noise, an initial condition $(Q_0,P_0)$ follows
a continuous random walk in the deterministic flow $\vec{K}$. The
probability density (PDF) $\rho(x,y)$ of reaching $(Q,P)$ from
some initial $(Q_0,P_0)$ satisfies the Fokker-Plank equation
(FPE): $\frac{\partial \rho}{\partial T} - \left\{g, \rho \right\}
= \vec{\nabla}\cdot\left(\Omega^{-1}\vec{r}\rho +
D\vec{\nabla}\rho\right)$. The solution can be written as a
functional integral of
$\exp{\left[-\frac{1}{D}\int_{(Q_0,P_0)}^{(Q,P)}
\mathcal{L}\left(Q'(T),P'(T)\right)\,dT\right]}$ over all paths
\cite{Mark '79},
where $\mathcal{L} = \frac{1}{4}\left[\left(\dot{Q} - K_Q\right)^2
+ \left(\dot{P} - K_P\right)^2\right]$. One can re-express all
paths as variations around the optimal path that minimizes the
exponent. Then $\rho(Q,P) =
\rho'(Q,P)\exp{\left[-\frac{1}{D}min\int_{(Q_0,P_0)}^{(Q,P)}
\mathcal{L}\left(Q'(t),P'(T)\right)\,dT\right]} =
\rho'(Q,P)\exp{\left[-S(Q,P)/D\right]}$ where the prefactor
$\rho'(Q,P)$ comes from performing path integrals over variations
around the optimal path. This pre-factor is \emph{nonexponential}
because the path integral over variations is to the lowest order a
Gaussian path integral. 
We are interested in the exponential factor, because in the limit
of weak noise it will dominate transition rates.

Now, the probability of undergoing a large fluctuation from a FP
to cross the separatrix is proportional to the probability of
being found in the most likely place where the exit traffic takes
place - the saddle point $(Q_s,P_s)$ \cite{Mark '79}. Therefore,
the rate of escape $W$ is to be approximated by $\rho(Q_s,P_s)$.
To calculate the probability of escape up to this accuracy, we
need in principle to solve Euler-Lagrange equations to find a path
that leads from the attractor to the saddle and evaluate the
action $S^*$ along this path. The Hamiltonian associated with
$\mathcal{L}$ is
\begin{equation}
\label{eq:H} \mathcal{H} = \left(p_Q^2 + p_P^2\right) + p_QK_Q +
p_PK_P
\end{equation}
$\mathcal{H}$ is often called the "auxiliary" Hamiltonian, to
avoid confusion with $g$.  The points
$(Q_{FP},P_{FP},P_Q=0,P_P=0)$ are fixed points of the auxiliary
dynamical system formed from $\mathcal{H}$.  Let's note that there
are infinitely many \emph{optimal} trajectories from attractor to
the saddle - they all differ by their initial momenta, but only
paths that have zero energy give the \emph{stationary}
$\rho(Q,P)$.  One way to see this is to substitute $\rho(Q,P) =
\rho'(Q,P)\exp{\left(-S(Q,P)/D\right)}$ into the \emph{stationary}
FPE, and obtain \cite{MS SIAM} $\left(\frac{\partial S}{\partial
Q}\right)^2 + \left(\frac{\partial S}{\partial P}\right)^2 +
\frac{\partial S}{\partial Q}K_Q + \frac{\partial S}{\partial
P}K_P$ = 0. This is just a Hamilton-Jacobi equation,
$\mathcal{H}(Q,P,\frac{\partial S}{\partial Q},\frac{\partial
S}{\partial P}) = 0$, and it describes dynamics on the
$\mathcal{H} = 0$ manifold.  Now, since $\vec{K}(Q_{FP},P_{FP}) =
0$, the momenta at all the FPs must also be zero.

In summary, in the limit of weak noise, the rate of escape $W
\approx \exp{\left(-R/D\right)}$ where $R=S^*$ - the action along
the trajectory which connects the attractor to the saddle and
satisfies $\mathcal{H} = 0$.

\subsection{Analytical Limits}
Both limits discussed below are physically analogous to the two
types of limits considered by Kramers \cite{Kramers}: an
overdamped limit and a low damping limit.  One major difference is
that our Hamiltonian $g(Q,P)$ does not have an explicit separation
of kinetic and potential energies.
\subsubsection{Overdamped regime in the vicinity of a bifurcation.}
\begin{figure}[h]
\begin{center}
\includegraphics[width=8.5cm]{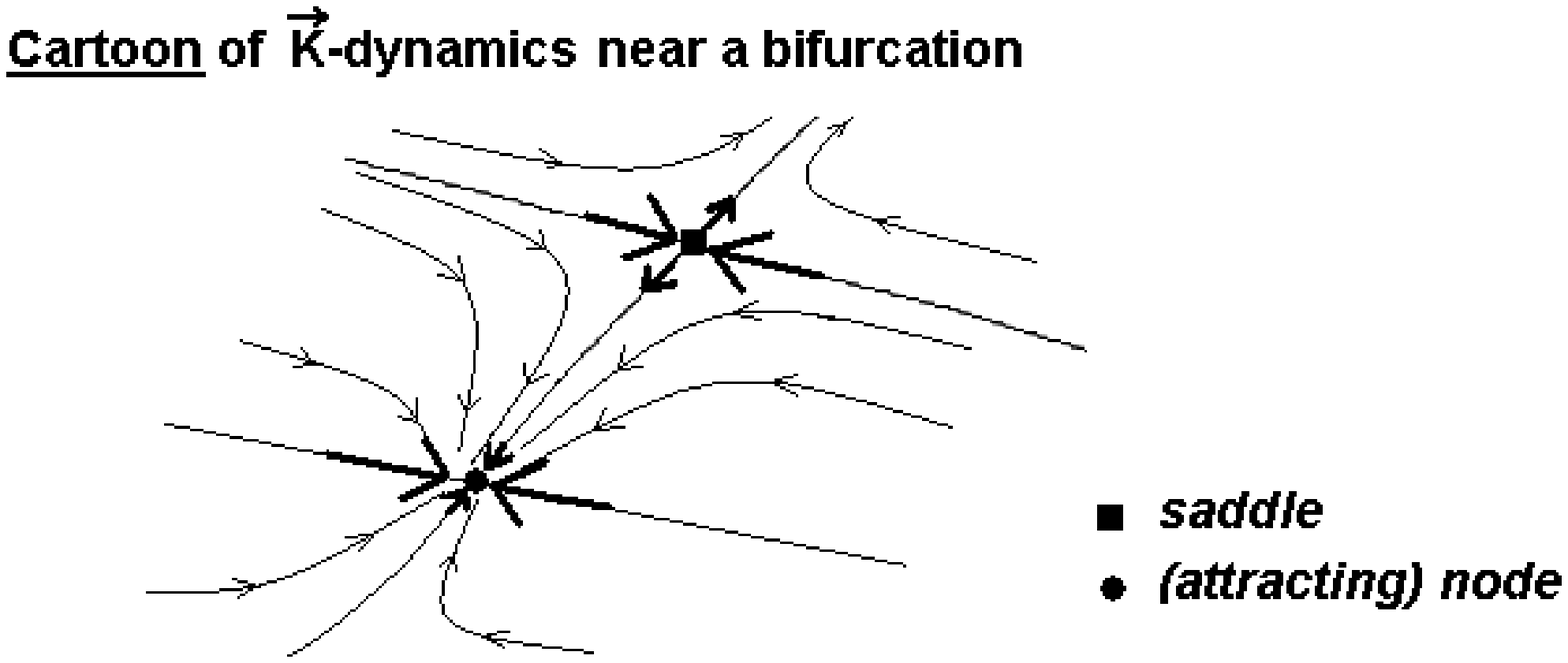}
\end{center}
\caption{\label{fig:Cartoon} Cartoon of the $\vec{K}$-flow near a
saddle-node bifurcation.}
\end{figure}

At either $\sqrt{\beta_B^{l}}$ or $\sqrt{\beta_B^{h}}$, the saddle
and one of the attracting nodes of $\vec{K}$ merge together at
$(Q_B,P_B)$ in a saddle-node bifurcation.  As $\sqrt{\beta}$
approaches $\sqrt{\beta_B}$, a saddle and a node approach each
other, the attractive eigenvector of the saddle, and one of the
attracting eigenvectors of the node align parallel to each other,
and the corresponding attracting eigenvalues of both FP approach
the same negative value.  Also, the repulsive eigenvector of the
saddle and the other attracting eigenvector of the node become
equal and the corresponding eigenvalues both tend to zero. This
process is depicted in the cartoon in Fig.~\ref{fig:Cartoon}.
\\
Thus, close to a bifurcation, there develops a soft mode - a
narrow region connecting the saddle with an attractor along which
the $\vec{K}$-motion is slow.  Due to this slowing down, the soft
mode forms a path of least resistance along which a large
noise-induced fluctuation away from the attractor is most likely
to take place.  To analyze this system we rewrite $\vec{K}$ in
terms of parameter $\tilde{\eta}$ and variables $(f = Q - Q_B, s =
P - P_B)$; it turns out to be unnecessary to make a linear
transformation to eigen-coordinates of $\vec{K}$ at $(Q_B,P_B,
\sqrt{\beta_B})$.  In terms of these new coordinates, $\vec{K}$
transforms to the following form: \small
\begin{eqnarray}
\dot{f} &=& -2\Omega^{-1}(f - a_B s) + P_Bf^2 +
2Q_Bsf + 3P_Bs^2 + f^2s + s^3 + n_Q \nonumber \\
\\
\dot{s} &=& \pm \tilde{\eta} - 3Q_B f^2 - 2P_Bfs - Q_Bs^2 - f^3 -
s^2f + n_P
\end{eqnarray}
\normalsize Here the $+$ sign applies for the low field (high
amplitude) bifurcation and
\begin{equation}
\label{eq:a_B-eqn} a_B = \Omega(1 \pm 2\sqrt{1-3\Omega^{-2}})/3.
\end{equation}
Notice that the position of fixed points sets characteristic
scaling of both $s$ and $f$ variables to be $\propto
\tilde{\eta}^{1/2}$. From this we see that $s$ is a slow variable
while $f$ is fast - on its time scale, $s$ appears approximately
frozen. To lowest order in $\tilde{\eta}$, we treat $s$ as
completely frozen, in which case $f$ relaxes to $a_Bs +
\mbox{corrections of order } \tilde{\eta}$. In such adiabatic
limit, the dynamics of $s$ is then given by
\begin{equation}
\dot{s} \approx \pm \tilde{\eta} - b(\Omega) s^2 + n_P
\end{equation}
where
\begin{equation}
b(\Omega) = 3Q_Ba_B^2 + 2P_Ba_B + Q_B
\end{equation}
($b>0$ for low field bifurcation and $b<0$ for high field
bifurcation). The potential for this slow coordinate is
\begin{equation}
U = \mp \tilde{\eta} s + \frac{b}{3}s^3,
\end{equation}
while the distance between the saddle and the attractor is
$\sqrt{\tilde{\eta}/|b|}$, thus the potential barrier for the slow
coordinate is
\begin{equation}
\Delta U = \frac{4\tilde{\eta}^{3/2}}{3|b|^{1/2}(\Omega)}.
\end{equation}
This is the well known $\xi=3/2$ scaling law.  A 1-dimensional FPE
for $s$ can be written, a stationary solution for
$\rho(s,\tilde{\eta},\Omega)$ obtained, and the escape rate
calculated. This quantity will have the form $W =
W'(\Omega)\exp{\left(-\Delta U/D\right)}$. Such adiabatic approach
is valid only sufficiently close to the bifurcation point.  When
$\Omega \ll 1$, "sufficiently close" means that $\tilde{\eta} \ll
\Omega^{-3}$. In this regime,
\begin{equation}
\label{eq:1d barrier} R \approx \left\{
\begin{array}{ll}
\frac{4\sqrt{2}}{3}\Omega^{-1/2}\tilde{\eta}^{3/2} & \mbox{low
field
bifurcation} \\
\frac{4}{3^{1/4}}\Omega^{-1}\tilde{\eta}^{3/2} & \mbox{high field
bifurcation}
\end{array} \right.
\end{equation}
A more systematic treatment of the close-to-a-bifurcation regime
which will allow us to extract corrections to the adiabatic limit
and derive the $\tilde{\eta} \ll \Omega^{-3}$ criterion will be
considered in Section III.A
\subsubsection{Low dissipation regime.}
When effective friction, $\Omega^{-1}$ is low, the dynamics of
$\vec{K}$ is approximately Hamiltonian: the time scale for the
decay of energy is much larger then the time scale to make one
cycle on the contour of approximately fixed energy.  This
separation of time scales breaks down very close to a bifurcation
point, so as long as $\Omega^{-1} \neq 0$, there will always be a
small fraction of the hysteresis displaying a $\xi=3/2$ scaling.
Aside from this very narrow region, the separation of time scales
allows one to turn a 2-variable FPE into a 1-variable FPE for the
diffusion of \emph{energy}, by averaging out the fast part of the
dynamics.  The noise strength $D$ is proportional to
$\Omega^{-2}$, so the FPE in the stationary regime can be
rewritten as $\left\{g, \rho \right\} = - \Omega^{-1}
\vec{\nabla}\cdot\left(\vec{r}\rho +
D' \Omega^{-1} \vec{\nabla}\rho\right)$. 
Note that when $\Omega^{-1}=0$, \emph{any} function of the form
$\rho(g(Q,P))$ will be a solution to this FPE.   Therefore, we can
expand $\rho(Q,P) = \rho^{(0)}(g(Q,P)) + \Omega^{-1}
\rho^{(1)}(Q,P) + O(\Omega^{-2})$ and integrate over a contour of
constant $g(Q,P)$.  We will refer to such theory as
"quasi-Hamiltonian" theory.  After some
algebra, 
this leads to:
\begin{equation}
\label{eq:FPE_en} \frac{d}{dE}\left(B(E) \rho^{(0)}(E) + D'
\Omega^{-1} D(E)\frac{d\rho^{(0)}}{dE} \right) = 0
\end{equation}
where $B(E) = \int \int_{g(Q,P) = E} \,dQ \,dP$ and $D(E) = \int
\int_{g(Q,P)=E} \nabla^2 g(Q,P) \,dQ \,dP$.  
An alternative approach followed by \cite{Mark and Ira} is to
calculate an average rate of energy decay due to friction and an
effective diffusion coefficient for the energy drift due to noise.
These would turn out to be precisely the $B(E)$ and $D(E)$
respectively, hence Eq.~(\ref{eq:FPE_en}) is the FPE for the
energy. Returning back from $D'$ to $D$, the activation rate is $W
\propto \exp{\left(-R/D\right)}$ where
\begin{equation}
\label{eq:quasi-ham theory}  R = -\Omega^{-1}\int_{E_{a}}^{E_s}
\frac{B(E')}{D(E')} \,dE'
\end{equation}
Here $E_a$ is the energy of the attractor, and $E_s$ is the energy
of the saddle. Close to bifurcation points, but not close enough
for the overdamped theory to be applicable,
\begin{equation}
\label{eq:quasi-ham theory assymp}
 R \approx \left\{
\begin{array}{ll}
2\Omega^{-1}\tilde{\eta} & \mbox{low field
bifurcation \cite{Mark '79}} \\
\frac{4}{3^{1/4}}\Omega^{-1}\tilde{\eta}^{3/2} & \mbox{high field
bifurcation \cite{Dmittriev and D'yakonov}}
\end{array} \right.
\end{equation}
Comparisson of Eqn. (\ref{eq:quasi-ham theory assymp}) with Eqn.
(\ref{eq:1d barrier}) reveals that there must be a crossover from
a $\xi = 1$ regime to $\xi = 3/2$ regime as $\sqrt{\beta}$
approaches $\sqrt{\beta_B^l}$.
%

\subsection{Power law scalings of escape barriers}
When $\Omega^{-1} \neq 0$, sufficiently close to the saddle-node
bifurcation, the dynamics near the attractor and the saddle is
generically over-damped \cite{Guckenheimer and Holmes}.  Hence,
the soft-mode geometric picture offered in Section II.A.1 is
generic and $\xi=3/2$ is a genetic property in a system with
finite damping and close enough to the saddle-node bifurcation.
Further from the bifurcation, the over-damped soft mode does not
exist. We have seen that for sufficiently low $\Omega^{-1}$, the
overdamped $\xi=3/2$ crosses over to $\xi=1$ (close to
$\sqrt{\beta_B^l}$) or remains $\xi=3/2$ (close to
$\sqrt{\beta_B^h}$). As recently proved by Dykman, Schwartz and
Shapiro \cite{Mark and Ira} (DSS), these are in fact also
\emph{generic} scaling laws in the under-damped vicinity of a
saddle-node bifurcation.
DSS considered two scenarios which may happen when $\Omega^{-1}$
\emph{equals} zero.  As $\tilde{\eta} \rightarrow 0$, the center
and the saddle may merge, or they may remain separate.  DSS named
the first type of a \emph{Hamiltonian} bifurcation "local" and the
latter "non-local". The two situations dictate the form of the
Hamiltonian $g$.

Let's switch to coordinates $(q,p)$ such that when $\Omega^{-1} =
0$, both the center and the saddle fixed points lie on the $p=0$
axis. In the case of a local Hamiltonian bifurcation, in the
vicinity of $(q=0, p=0, \tilde{\eta} = 0)$, $g$ is given by
$g_{L}(q,p,\tilde{\eta}) = \frac{1}{2}p^2 + U_{cub}(q)$ where
$U_{cub}(q) = -\frac{1}{3}q^3 + \tilde{\eta} q$.  The center lies
inside a small homoclinic loop at the energy of the saddle. DSS
proved that given this form of the Hamiltonian $g$, addition of
noise and damping leads to $R(\tilde{\eta}) = A(\Omega)
\tilde{\eta}^{3/2}$ (in the under-damped regime).  But we know
that as as the system approaches the bifurcation even closer, it
becomes over-damped, a soft-mode sets up and the method of
analysis presented in Section II.A.1 leads to $R(\tilde{\eta}) =
B(\Omega) \tilde{\eta}^{3/2}$. Therefore, assuming smooth
dependence upon $\tilde{\eta}$, it must be that $A(\Omega) =
B(\Omega)$. In other words, the $\xi = 3/2$ power law in this
situation holds beyond where it was expected to hold due to a soft
mode theory!  In the Duffing system, a nonlocal bifurcation takes
place at $\sqrt{\beta_B^h}$.  In the numerical experiments
described below, we found that the $\xi=3/2$ power law holds near
$\sqrt{\beta_B^h}$ over a \emph{finite} range of $\tilde{\eta}$
even as $\Omega^{-1} \rightarrow 0$. This $\xi=3/2$ law at zero
$\Omega^{-1}$ has been predicted by \cite{Dmittriev and
D'yakonov}, but the new work of DSS gives it a new explanation. We
elaborate upon this in Section III.B.

In the case of a nonlocal Hamiltonian bifurcation, $g$ is given by
$g_{NL}(q,p,\tilde{\eta}) = \frac{1}{2}p^2 + \tilde{\eta} U(q)$,
where $U(q)$ is independent of $\tilde{\eta}$ and has a local
minimum and maximum.  The homoclinic loop of such systems has the
property that only its size in the $p$-direction shrinks to zero
as $\tilde{\eta} \rightarrow 0$, while the distance between the
center and the saddle inside the loop remains finite. Introduction
of damping into these types of systems will not change the
separation between the saddle and the attractor until the system
has come close enough to the bifurcation: $\tilde{\eta} \sim
\Omega^{-2}$. DSS proved that given this form of the Hamiltonian
$g$, addition of noise and damping leads $R(\tilde{\eta}) \propto
\tilde{\eta}$ (in the underdamped regime). Therefore, at finite
$\Omega^{-1}$ we expect a $\xi = 1$ scaling moderately close to
the bifurcation, but further away then the region of over-damped
$\xi=3/2$ scaling.  In the Duffing system, a nonlocal bifurcation
takes place at $\sqrt{\beta_B^l}$.
It is important to stress that all saddle-node bifurcations at
\emph{finite} $\Omega^{-1}$ are local in the sense that the
attractor and the saddle merge as $\tilde{\eta} \rightarrow 0$.
The terms "local" or "nonlocal" only make sense at zero
$\Omega^{-1}$. We must also mention that a nonlocal bifurcations
may take place in many forms. For example, in a Duffing
oscillator, the homoclinic loop has a horse-shoe shape at
$\Omega^{-1} = 0$ which becomes thinner while remaining extended
as $\tilde{\eta} \rightarrow 0$. DSS found a change of variables
which maps from $g$ to $g_{NL}$.

\section{Numerical Method and Results}
Numerical calculations of escape rate are based on direct
evaluation of $S^*$. The algorithm for doing so is as follows. (1)
Choose parameters $\Omega^{-1}$ and $\sqrt{\beta}$. (2) Calculate
repulsive eigenvectors (i.e. those whose eigenvalues have positive
real part) of the FP $(Q_{attr},P_{attr},p_Q=0,p_P=0)$ of the
auxiliary system.  These eigenvectors span the tangent plane to
the unstable manifold of this attractor. (3) Position a locus of
initial conditions on a small circle that lies on this tangent
plane and centered around this FP and evolve each of these initial
conditions according Hamiltonian equations with Hamiltonian
$\mathcal{H}$ from Eqn. (\ref{eq:H}). The angle around this
circle, $\varphi$, serves as a parameter that enumerates a
trajectory. (4) From this set of trajectories, find the one which
leads into the saddle, with a special trajectory parameter
$\varphi^*$. Because the method involves "shooting" many
trajectories to see which one hits the saddle, this method is
occasionally called "the shooting method". (5) Evaluate the action
along this trajectory. For each $\Omega^{-1}$, this procedure
would be repeated for many values of $\sqrt{\beta}$ between
$\sqrt{\beta_B^l}$ and $\sqrt{\beta_B^h}$.

In practice, step (4) may require several rounds of bracketing. In
each round, for each trajectory, we calculate the distance between
the saddle and the point of intersection of that trajectory with
the separatrix, measured along the separatrix.  We identify a
$\varphi$ for which the trajectory came closest to the saddle and
then evolve another set of trajectories in a small range of
trajectory parameters around this $\varphi$. After more and more
rounds of such procedure, we would obtain a trajectory that hits
the separatrix closer and closer to the saddle.  In general,
finding a trajectory that connects an attractor with the saddle is
difficult because it lies on the intersection of the unstable
manifold of the attractor and the stable manifold of the saddle in
the 4D auxiliary space \cite{MS SIAM}, \footnote{The question of
\emph{existence} of this intersection, to our knowledge, is not a
fully understood one.}.  A small deviation of $\varphi$ on
opposite side of $\varphi^*$ will produce trajectories that lie on
opposite sides of this intersection and will diverge from each
other.  This effect is most pronounced closer to bifurcations
because the motion along the optimal trajectory becomes much
slower then the diverging motion away from it - such is the
structure of eigenvalues of both attractor and saddle close to the
bifurcation. This polarity takes place only close to bifurcation
points, but it persists further away from bifurcation points when
$\Omega^{-1}$ is larger. Therefore, in the small-$\Omega^{-1}$
regime, finding a trajectory between an attractor and the saddle
is difficult only very close to bifurcation points, but in the
large-$\Omega^{-1}$ regime, finding such a trajectory is difficult
over a larger part of the hysteresis.

We will now summarize results of the computation of $S^*$, scaling
laws, and crossovers. It is sometimes easier to present results in
terms of $x = \frac{\tilde{\eta}}{\sqrt{\beta_B^r} -
\sqrt{\beta_B^l}}$ - "reduced" $\tilde{\eta}$. For each
$\Omega^{-1}$ and each type of transition, we scanned the
hysteretic region at multiple values of $x$ and for each $x$ we
followed the "shooting" procedure, and where necessary, the
bracketing method.
\subsection{Up $\rightarrow$ Down transitions}
\subsubsection{Numerical results}
\begin{figure}[h]
\includegraphics[width=8.5cm]{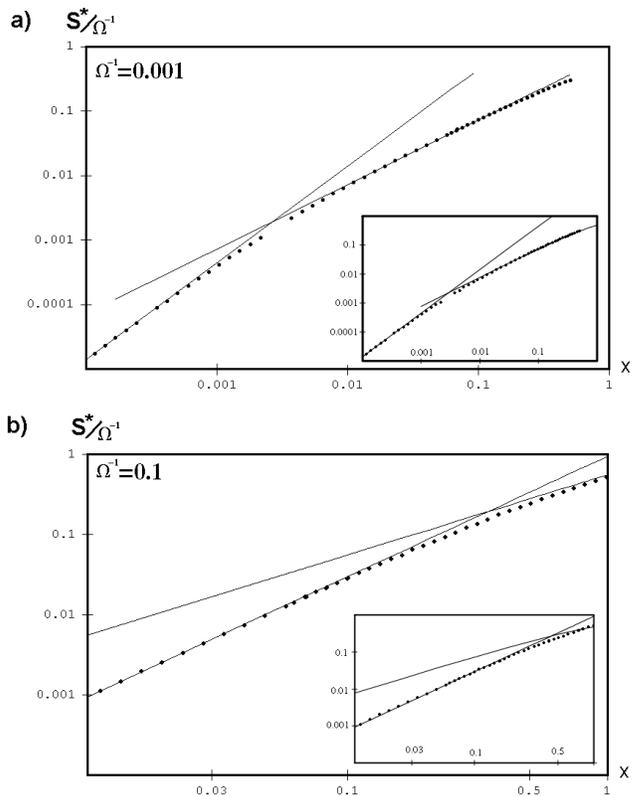}
\caption{\label{fig:up-down crossover} Main plots:
$S^*(x)/\Omega^{-1}$ (dots) and \emph{fits} (lines) for up
$\rightarrow$ down transitions. The fit to the $\xi=3/2$ regime
was made by analyzing $\ln{\left(S^*/\Omega^{-1}\right)}$ vs. $
\ln{x}$ and a fit to the $\xi=1$ regime was made by analyzing
$S^*/\Omega^{-1}$ vs. $x$. Inserts: $S^*(x)/\Omega^{-1}$ (dots)
and \emph{theory} (lines) based on Eqn. \ref{eq:1d barrier} for
the $\xi=3/2$ regime and Eqn. \ref{eq:quasi-ham theory assymp} for
the $\xi=1$ regime.}
\end{figure}

\begin{figure}[h]
\begin{center}
\includegraphics[width=8.5cm]{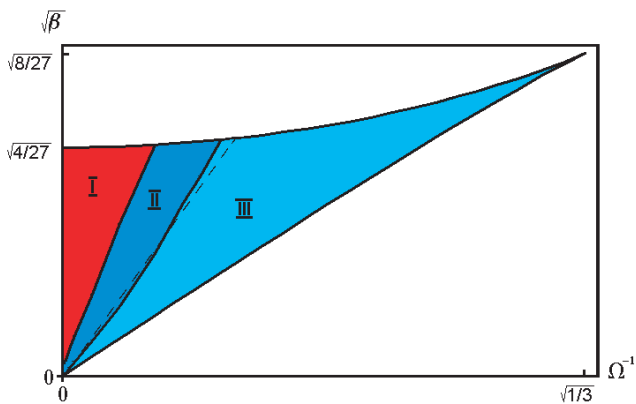}
\end{center}
\caption{\label{fig:Phase diagram up-down} Regions of different
scaling behaviors for up $\rightarrow$ down transitions obtained
by scanning $\sqrt{\beta}$ at fixed $\Omega^{-1}$ at multiple
values of $\Omega^{-1}$. Region I - non-power-law regime.  Region
II - $\xi=1$ regime and region III - $\xi=3/2$ regime. The
crossover between the $\xi=1$ and a non-power law regime was
defined as the point at which the fit to $\xi=1$ regime came
closest to the $\ln{\left(Q^*/\Omega^{-1}\right)}$ vs. $\ln{x}$ -
 see Fig.~\ref{fig:up-down crossover}.}
\end{figure}

\noindent A plot of $S^*/\Omega^{-1}$ versus $x$, Figs.
\ref{fig:up-down crossover}, \ref{fig:Linear scale} reveals
several features. At low $\Omega^{-1}$ two regimes clearly stand
out: one with $\xi \approx 3/2$, and one with $\xi \approx 1$.
Further from the bifurcation, the $\xi=1$ scaling law breaks down
and $S^*/\Omega^{-1}$ versus $x$ displays a non-power-law
behavior. For various values of $\Omega^{-1}$, the data in the
"$3/2$
regime" fits well to a power law with $\xi$ which is up to $3\%$
below the theoretical value of $3/2$. In general, \emph{precise}
extraction of the exponent $3/2$ requires approaching very close
to the bifurcation. It appears that the discrepancy was highest in
those cases when the bifurcation was approached less closely
(relative to the crossover point) or less reliably. The
reliability of the approach is hampered at larger $\Omega^{-1}$ by
the strong divergence of trajectories explained in the previous
section. A crossover between the $\xi=3/2$ and $\xi=1$ regime is
defined to be such $x$ at which fits to the respective regions
intersect. Collecting data for $S^*$ vs. $x$ for various
$\Omega^{-1}$ we were able to map out the entire bi-stability
region for locations of various scaling regimes and crossovers
between them. The lowest $\Omega^{-1}$ at which computations were
made was $10^{-3}$. The product of this work is depicted in
Fig.~\ref{fig:Phase diagram up-down}.  It is important to stress
that in Fig.~\ref{fig:Phase diagram up-down} for up $\rightarrow$
down transitions and in Fig.~\ref{fig:Phase diagram down-up} for
down $\rightarrow$ up transitions, the mapping was performed by
scanning $\sqrt{\beta}$ while holding $\Omega^{-1}$ fixed.
\\
Note that as $\Omega^{-1} \rightarrow 0$, the linear scaling is
found between $x = 0$ and some finite $x$.  At small, but non-zero
$\Omega^{-1}$, the region of $\xi=3/2$ subsequently grows.  These
results are in accord with the theorem of DSS which states that
close to a bifurcation which at $\Omega^{-1}=0$ becomes nonlocal,
in the underdamped regime at non-zero $\Omega^{-1}$ the $\xi$ is
expected to be $1$.  Since the underdamped portion of the
hysteresis extends all the way to the bifurcation as $\Omega^{-1}
\rightarrow 0$, we expect the linear scaling to extend all the way
to the bifurcation in this limit \footnote{In practice we could go
down to $\Omega^{-1} = 0.001$.  In principle, strictly at
$\Omega^{-1} = 0$ escape trajectories do not exist; they only have
meaning in the limit $\Omega^{-1} \rightarrow 0$.}.

Clearly, the boundaries between regions I, II and III are purely
conventional - they have been defined in some convenient way, but
do not correspond to anything specifically physical.  To emphasize
this point, it helps to construct a 3-dimensional plot of
$S^*/\Omega^{-1}$ versus $x$ and $\Omega^{-1}$ - Fig.~\ref{fig:3D
up-down}.
\begin{figure}[h]
\begin{center}
\includegraphics[width=8cm]{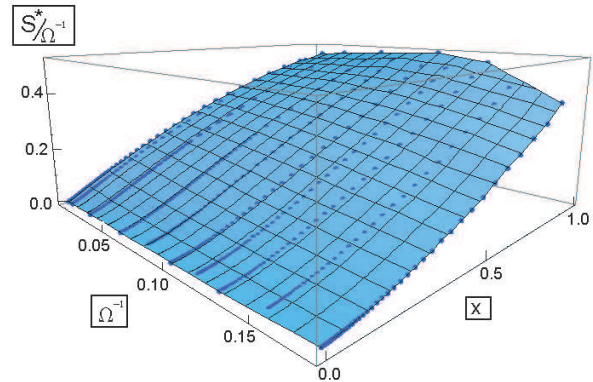}
\end{center}
\caption{\label{fig:3D up-down} $S^*/\Omega^{-1}$ versus $x$ and
$\Omega^{-1}$ up to $\Omega^{-1} = 0.175$ which is about $30\%$ of
the hysteresis.  Fig.~\ref {fig:up-down crossover} are two slices
of this surface.}
\end{figure}
This surface is smooth and large parts of it happen to fit well to
power laws.  The regions when this is possible appear to be much
bigger then the regions at which the $\xi = 1$ and $\xi = 3/2$
scaling was supposed to be applicable on the basis of
approximations discussed in Section II.  A natural question is why
this is so!  To shed more light on this issue, let's digress on
the crossover from the $\xi=3/2$ to $\xi = 1$ in more detail.
\subsubsection{Discussion of the $\xi=3/2$ to $\xi=1$ crossover}
Notice that the crossover between the $\xi=3/2$ and $\xi=1$
regime, as it appears in Fig.~\ref{fig:up-down crossover} was
defined as the point of intersection of the fits to the respective
regimes.  Thus, to predict this crossover we simply equate the
asymptotic in the $\xi=3/2$ regime, $R =
\frac{4\sqrt{2}}{3}\Omega^{-1/2}\tilde{\eta}^{3/2}$ to the
asymptotic in the $\xi=1$ regime $R=2\Omega^{-1}\tilde{\eta}$
(this is most accurate as $\Omega^{-1} \rightarrow 0$).  The
result is $\tilde{\eta} = \frac{9}{8}\Omega^{-1}$, which
corresponds to $\sqrt{\beta} = \sqrt{\beta_B^l} +
\frac{9}{8}\Omega^{-1}$ $\left(\sqrt{\beta_B^l} \mbox{is itself a
function of } \Omega^{-1}\right)$.  The cross-over predicted this
way is depicted by dashed line in Fig.~5. The fact that this
method of predicting a crossover works well tells us that the
asymptotic expressions are correct.  However, we could imagine
that the width of the crossover region was very large, in which
case the crossover defined by equating the fits and predicted by
equating the asymptotics is not physically meaningful.  The width
of the crossover region is a more physically interesting quantity.
Alternatively we may also define the following two crossovers: a
"lower crossover" (LC) defined as a characteristic value of
$\sqrt{\beta}$ (or $\tilde{\eta}$ or $x$) above which the
$\xi=3/2$ regime begins to fail and an "upper crossover" (UC)
defined as a characteristic value of $\sqrt{\beta}$ (or
$\tilde{\eta}$ or $x$) below which the $\xi=1$ regime begins to
fail. The width of the crossover region is the difference between
these two values.

Physically, we expect the LC to take place where the soft-mode
picture fails.  Therefore, a good \emph{characteristic} estimate
of this will be such $\tilde{\eta}$ at which the real and
imaginary parts of eigenvalues become equal. This can be shown to
happen for $\tilde{\eta} = \Omega^{-3}/2$ - the scaling is
\emph{cubic}.  To predict the LC more accurately let's turn back
to the type of analysis described in section II.A.1. Let's
re-write Eqns. (5)-(6) for the non-adiabatic part of the fast
variable, $A = f - a_B s$. Near low-field bifurcation, these will
have the form
\begin{widetext}
\begin{eqnarray}
\dot{A} &=& - a_B \tilde{\eta} - 2\Omega^{-1} A + a_1 s^2 + a_2 As
+ a_3 A^2 + a_4 s^3 + a_5 s^2 A + a_6 sA^2 + a_7 A^3 + n_x - a_B n_y \\
\dot{s} &=& \tilde{\eta} + b_1s^2 + b_2 As + b_3 A^2 + b_4 s^3 +
b_5As^2 + b_6 A^2s + b_7 A^3 + n_y
\end{eqnarray}
\end{widetext}
All coefficients $a_i$ and $b_i$ are functions of $\Omega^{-1}$.
The position of fixed points sets the characteristic scaling of
$A$ to be $\tilde{\eta}$ and of $s$ to be $\tilde{\eta}^{1/2}$
(generic feature of saddle-node bifurcations), so we define new
variables: $s = \tilde{\eta}^{1/2}s'$ and $A = \tilde{\eta} A'$.
Moreover, the $\Omega$-dependencies of the coefficients $a_i$ and
$b_i$ near low-field bifurcation are such that when $\Omega^{-1}$
is small compared to $1$, further re-scaling of $s'$ and $A'$ as
$s' = \Omega^{-1/2}\sigma$ and $A' = \Omega^{2}\phi$ yields a
single dimensionless parameter $\mu = \tilde{\eta}\Omega^3$; such
reduction is approximate, but becomes more accurate for smaller
$\Omega^{-1}$ \footnote{From our definitions, $\omega_F = \omega_0
\left(1 + \frac{1}{Q\Omega^{-1}}\right)$. Recall that in order for
equations (\ref{eq:2D_dynamics1})-(\ref{eq:2D_dynamics2}) to
apply, we must be working in the regime when $\omega_F = \omega_0
(1 + \mbox{small number})$. Therefore, if the limit $\Omega^{-1}
\rightarrow 0$ is taken, it is implied that the limit $Q
\rightarrow \infty$ is taken first.}. In terms of $\mu$, $\sigma$
and $\phi$ we have
\begin{widetext}
\begin{eqnarray}
\label{eq:long1} \frac{d\phi}{d\tau} &=& \left(-1-2\phi +
\frac{3}{2}\sigma^2\right) + \mu^{1/2} \left(\sigma\phi + \sigma^3
\right)
+ \mu \left(3\phi\sigma^2 - \frac{\phi^2}{2}\right) + \mu^{3/2} 3\sigma \phi^2 + \mu^2 \phi^3 + N_{\phi} \\
\label{eq:long2} \frac{d\sigma}{d\tau} &=& \mu^{1/2} \left(1 -
\frac{\sigma^2}{2}\right) + \mu\left(\sigma \phi - \sigma^3\right)
+ \mu^{3/2} \left(\frac{3}{2}\phi^2 - 3\sigma^2\phi \right) -
\mu^2 3\sigma \phi^2 - \mu^{5/2} \phi^3 + N_{\sigma}
\end{eqnarray}
\end{widetext}
where $\tau = \Omega^{-1}T$. 
With the help of the definition of $\sigma$ and
Eq.~(\ref{eq:a_B-eqn}) we note that the zeroth-order part of the
fast variable $a_Bs$ is given by $\tilde{\eta}^{1/2}
\Omega^{1/2}\sigma$ and the correction due to non-adiabaticity of
the fast variable is $A =
\tilde{\eta}\Omega^2\phi$.  We also see from
Eqs.~(\ref{eq:long1})-(\ref{eq:long2}) that when $\mu \ll 1$, both
$\sigma$ and $\phi$ are $O(1)$.  Hence comparison of the two terms
says that when $\mu \ll 1$, \emph{non-adiabatic part/adiabatic
part} $\sim
(\tilde{\eta}\Omega^2)/(\tilde{\eta}^{1/2}\Omega^{1/2}) =
\mu^{1/2} \rightarrow 0$ as $\mu \rightarrow 0$. In other words,
in this regime the non-adiabatic part becomes unimportant and the
problem reduces to 1 dimension defined by the noisy dynamics of
the slow variable.  To do this reduction explicitly, we first
notice from Eqs.~(\ref{eq:long1})-(\ref{eq:long2}) that at lowest
order, the variable $\sigma$ can be considered constant while
$\phi(\sigma) \approx \frac{3}{4}\sigma^2 - \frac{1}{2} +
\mbox{non-adiabatic corrections}$. This $\phi(\sigma)$ is then
substituted into Eq.~(\ref{eq:long2}).  
The resulting expression can only be considered up to $O(\mu)$
since the higher-order terms must also include the non-adiabatic
corrections to $\phi$ to be complete [note that if the
$\mu^{1/2}$-terms in Eq.~(\ref{eq:long2}) contained any $\phi$
terms, the complete expression within the adiabatic approximation
would only extend to $O(\mu^{1/2})$, not $O(\mu)$].  The barrier
height in the resultant 1-dimensional potential is $\Delta U =
\frac{4\sqrt{2}}{3}\mu^{1/2} + \sqrt{2}\mu^{3/2}$. Again, the
correction is within the adiabatic approximation; a non-adiabatic
correction would be higher order in $\mu$.  Taking into account that the 
$N_{\sigma}$ has a diffusion constant 
$D_{\sigma} = \frac{\Omega^2}{\tilde{\eta}}D$, 
we have
\begin{equation}
\label{eq:R-higher_order} R =
\frac{4\sqrt{2}}{3}\Omega^{-1/2}\tilde{\eta}^{3/2} +
\sqrt{2}\Omega^{5/2}\tilde{\eta}^{5/2}
\end{equation}
[compare with Eq.~(\ref{eq:1d barrier})].  We see again that the
correction term grows to be dominant when $\mu \sim 1$ (i.e. when
$\tilde{\eta} \sim \Omega^{-3}$) , but this is precisely when the
non-adiabatic part becomes comparable to the adiabatic part, and
it turns out to be meaningless to treat the
problem as effectively 1-dimensional.  
Our analysis taught us that in the regime when $\tilde{\eta} \ll
\Omega^{-3}$, the activation barrier $R$ will scale as
$\tilde{\eta}^{3/2}$ to a good approximation.  Outside the regime
of $\tilde{\eta} \ll \Omega^{-3}$ the current analysis only allows
us to conclude that if $\xi$ remains to be $3/2$, the physics
behind this scaling is not a physics of a 1-dimensional
over-damped soft mode; the escape problem is not 1-dimensional for
$\tilde{\eta} \gg \Omega^{-3}$.

To answer the question of what happens outside of the
$\tilde{\eta} \gg \Omega^{-3}$ regime we chose to compute the most
likely escape path numerically and calculate $S^*$ on that path.
We looked at the difference between the zeroth-order theory, $R_0
= \frac{4\sqrt{2}}{3}\Omega^{-1/2}\tilde{\eta}^{3/2}$ and the
numerically-computed $S^*$.
This was repeated for several values of $\Omega$ and in each case,
a fit to the function $\propto \tilde{\eta}^{p}$ was made.
Although this represents computer calculations, the data is
somewhat noisy due to the difficulty of hitting the saddle. Values
of $p$ between $2.15$ and $2.86$ were found.  The predicted value
of $p=5/2$ in the correction factor lies in the middle of this
range. However, the crossover LC, defined as an intersection of
some small fraction $r$ of $R_0$ with a fit to $R_1$ scales
\emph{linearly} with $\Omega^{-1}$, not cubically.
The $\tilde{\eta}_{LC}$ so obtained can be fit to a power law with
exponent $1.2$ for $r=0.1$ and  and $1.1$ for $r=0.2$ - both of
these exponents are far from $3$.

\begin{figure}[h]
\begin{center}
\includegraphics[width=7cm]{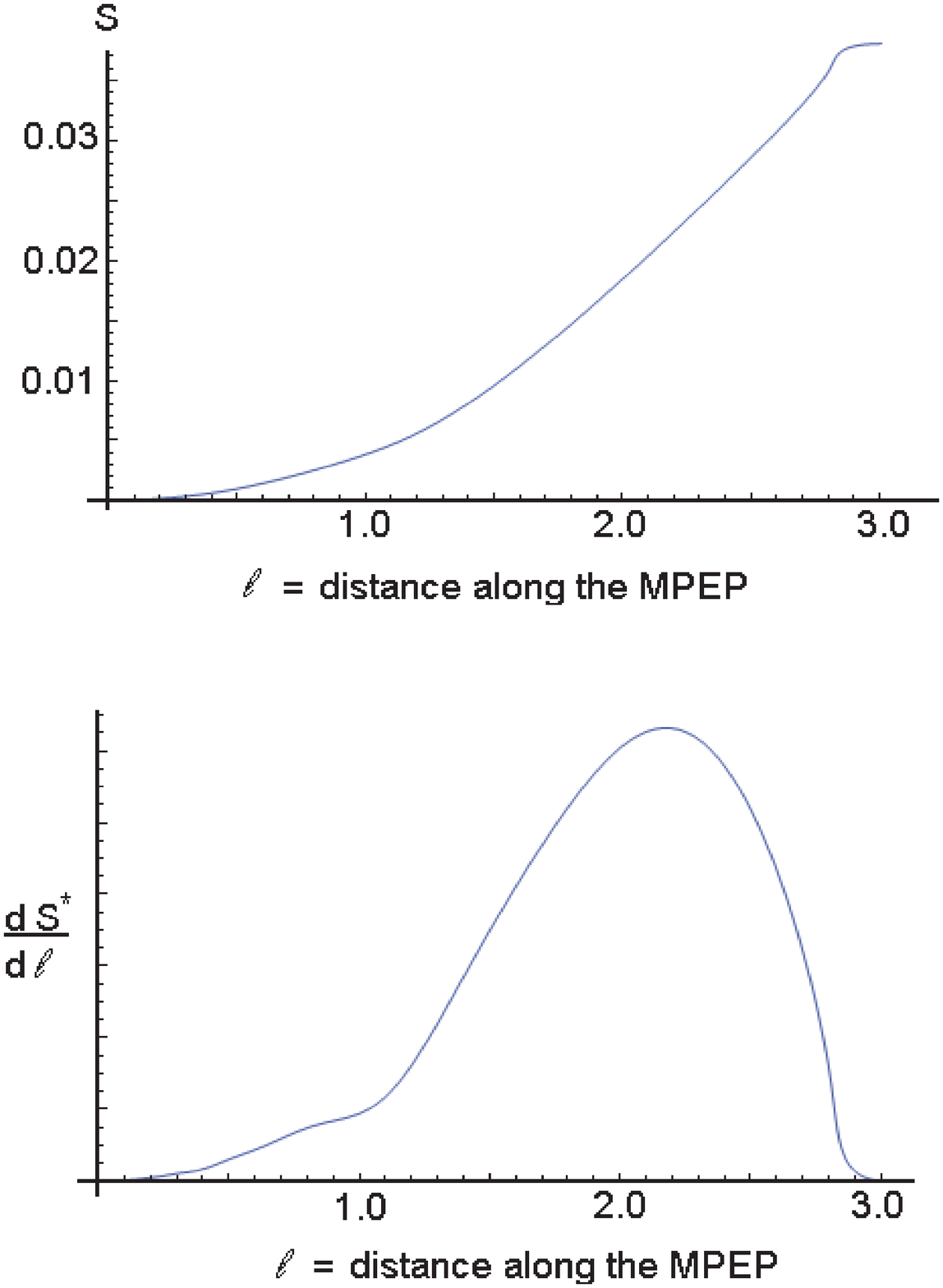}
\includegraphics[width=8cm]{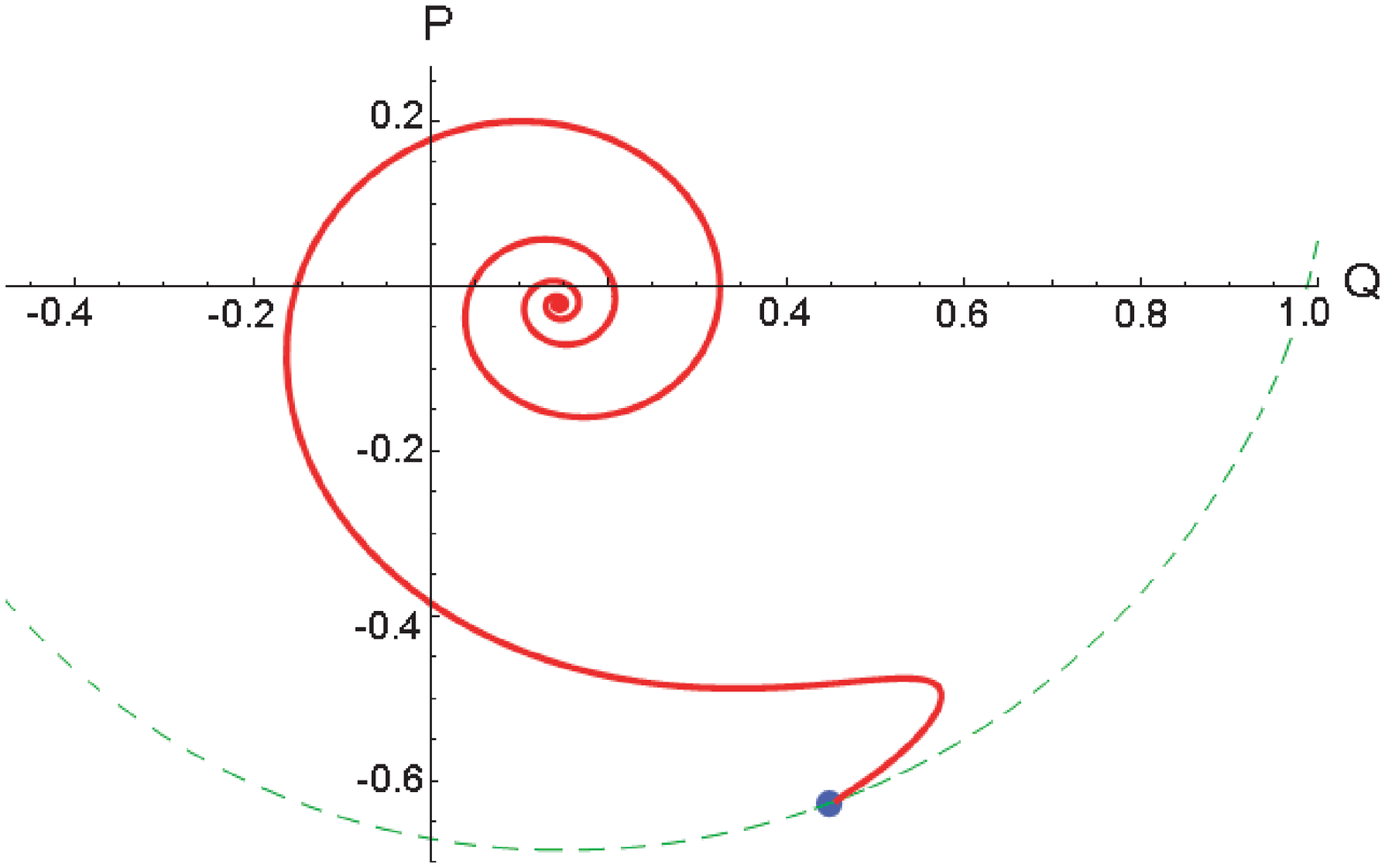}
\end{center}
\caption{\label{fig:Action on MPEP} $S(l)$ and its derivative for
particular parameters in region III as well as the plot of the
MPEP (dashed curve denotes separatrix). Here $\Omega^{-1} = 0.15$
and $\sqrt{\beta} = 10\%$ of the hysteresis. These parameters
correspond to $\mu \approx 5$; the soft-mode picture is expected
to hold only for $\mu \ll 1$.}
\end{figure}

The question of why the $\xi = 3/2$ scaling holds over a much
larger region of parameter space then expected based on the
1-dimensional soft-mode picture remains unanswered, and forms a
good challenge problem for a future work.  One potential
hypothesis is that in the region when the 1-dimensional soft-mode
picture does not hold true, the action along the MPEP actually
grows only over a small portion of the MPEP where the motion is
essentially 1-dimensional.  We provide plot of the action versus
the distance along the trajectory, $S(\ell)$ and the derivative
$dS/d\ell$ for a particular value of parameters deep in region III
- Fig.~\ref{fig:Action on MPEP}. The derivative plot clearly shows
that action grows fastest in the last winding of the trajectory,
but this is not sufficient to conclude that the dynamics in that
part of the trajectory can somehow be described as effectively
1-dimensional.

\begin{figure}[h]
\begin{center}
\includegraphics[width=8.4cm]{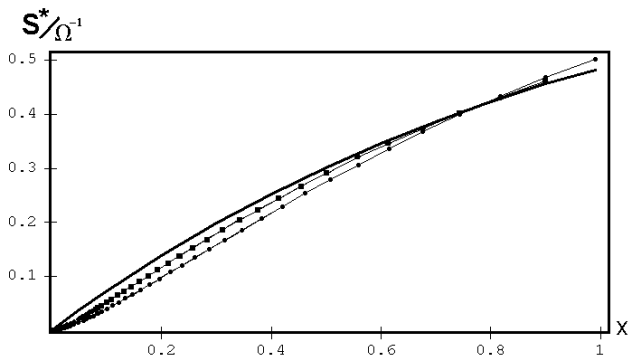}
\end{center}
\caption{\label{fig:Linear scale} $R^*(x)/\Omega^{-1}$ on a linear
scale. The thick black line represents the zeroth-order theory,
Eq.~(\ref{eq:quasi-ham theory}).  The other curves represent
numerical calculations of $S^*$ - squares: $\Omega^{-1} = 0.03$,
circles: $\Omega^{-1} = 0.06$.}
\end{figure}

Similarly to the LC, the UC can be defined as a characteristic
value of the driving field \emph{below} which the linear scaling
begins to break down. To predict this point, we must again compare
the lowest and one higher order terms in the theory for
$R(\tilde{\eta})$ that applies in the $\xi=1$ regime. The effect
of a non-zero $\Omega^{-1}$ is clearly seen in
Fig.~\ref{fig:Linear scale}
The lowest-order in dissipation theory (quasi-Hamiltonian) indeed
predicts linear scaling for low values of $\tilde{\eta}$ (or $x$),
which according to Eq.~(\ref{eq:quasi-ham theory assymp}) is given
by $2\Omega^{-1} \tilde{\eta}$.  The region of linear scaling
persists for non-zero $\Omega^{-1}$, but due to the existence of
the $3/2$-scaling region, this linear part has been "pushed over"
to larger $\tilde{\eta}$ and to lower values then those predicted
by the zeroth-order theory.  A work by Chinarov' et. al.
\cite{Chinarov} worked out dissipative corrections to the
quasi-Hamiltonian theory to next order in $\Omega^{-1}$.
Translating their results to our notation reads: $R =
\Omega^{-1}\left(2 \tilde{\eta} - \pi \Omega^{-1}\right)$ (compare
with Eq.~\ref{eq:quasi-ham theory assymp}).  The factor
$\pi\Omega^{-2}$ is just that correction which takes this lowering
into account.  
We expect the linear scaling to start breaking down when the
correction reaches some fraction $r$ of the zeroth-order term,
i.e. when $\tilde{\eta}_{UC} = \frac{\pi}{2r}\Omega^{-1} \approx
\frac{\pi}{2r}\Omega^{-1}$ for low $\Omega^{-1}$.  To extract
$\tilde{\eta}_{UC}$ from the data we subtracted the zeroth-order
part from $S^{*}/\Omega^{-1}$.  The resulting function grows for
low $\tilde{\eta}$, slows down for intermediate $\tilde{\eta}$
(but not completely because the exact slope in the linear regime
is slightly different from the zeroth-order result) and increases
for yet larger $\tilde{\eta}$ past the linear regime (see Fig.~6).
The average value in the intermediate regime is taken to be the
offset factor.  Equating this correction to the zeroth order term
yields $\tilde{\eta}_{UC}$ which does depend linearly on
$\Omega^{-1}$.
\subsection{Down $\rightarrow$ Up transitions}
\begin{figure}[h]
\begin{center}
\includegraphics[width=8.5cm]{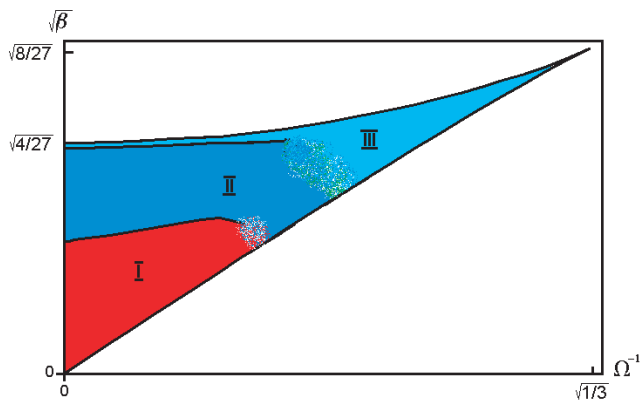}
\end{center}
\caption{\label{fig:Phase diagram down-up} Regions of different
scaling behaviors for down $\rightarrow$ up transitions obtained
by scanning $\sqrt{\beta}$ at fixed $\Omega^{-1}$ at multiple
values of $\Omega^{-1}$.  Region I corresponds to a non-power-law
regime. Region II corresponds to the $\xi \approx 1.3$ regime.
Region III corresponds to the $\xi=3/2$ regime.}
\end{figure}

Analogously to the up $\rightarrow$ down transitions, the entire
bi-stability region was mapped out. The product of this mapping is
shown in Fig.~\ref{fig:Phase diagram down-up}.  The lowest
$\Omega^{-1}$ at which computations were made was $0.005$.
\begin{figure}[h]
\begin{center}
\includegraphics[width=8cm]{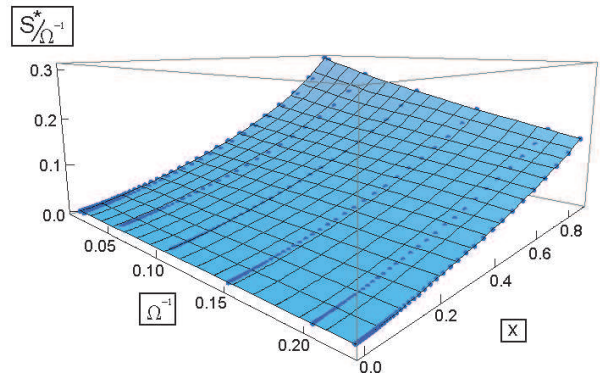}
\end{center}
\caption{\label{fig:3D down-up} $S^*/\Omega^{-1}$ versus $x$ and
$\Omega^{-1}$ up to $\Omega^{-1} = 0.22$ which is about $33\%$ of
the hysteresis.  Fig.~\ref {fig:up-down crossover} are two slices
of this surface.}
\end{figure}
Fuzzy regions correspond to those sections of the $\sqrt{\beta}$ -
$ \Omega^{-1}$ plane which proved exceptionally difficult to map
reliably due to high divergence of trajectories. A portion of the
surface $S^*/\Omega^{-1}(x,\Omega^{-1})$ is presented in
Fig.~\ref{fig:3D down-up}.

\begin{figure}[h]
\begin{center}
\includegraphics[width=8.6cm]{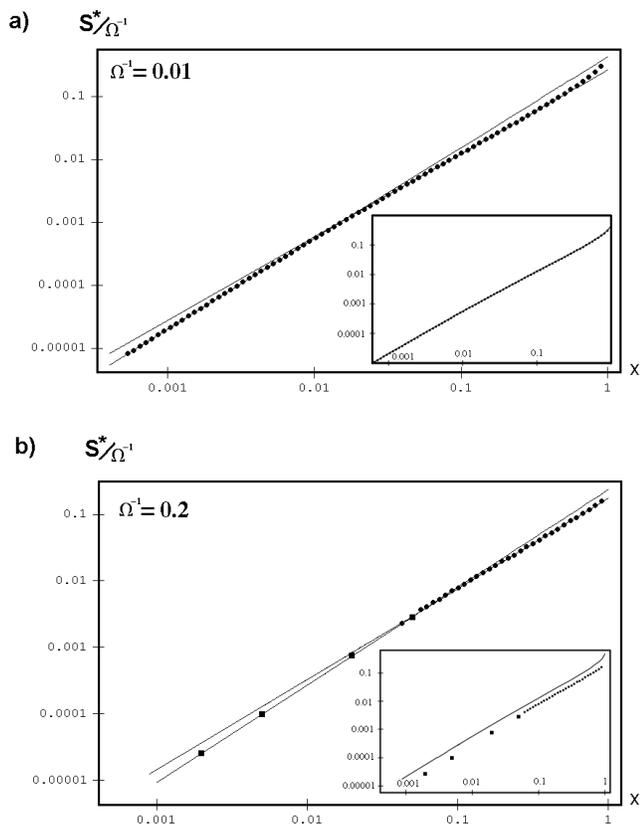}
\end{center}
\caption{\label{fig:down-up crossover} Main plots:
$S^*(x)/\Omega^{-1}$ (dots) and \emph{fits} (lines) for down
$\rightarrow$ up transitions.  Fits to the $\xi=3/2$ and $\xi
\approx 1.3$ regimes was made by analyzing
$\ln{\left(S^*/\Omega^{-1}\right)}$ vs. $ \ln{x}$. The region
where the simple shooting method was unreliable for hitting the
saddle (due to high divergence of trajectories), a bracketing
method was used, the results of which are denoted by squares.
Inserts: $S^*(x)/\Omega^{-1}$ (dots) and \emph{theory} (lines)
based on Eqn. \ref{eq:quasi-ham theory} for the $\xi=1$ regime;
note that this quasi-Hamiltonian theory predicts the $\xi=3/2$
scaling, as was first pointed out by \cite{Dmittriev and
D'yakonov} and in light of the recent work by \cite{Mark and Ira}
this fact has been explained to be a result of the fact that this
bifurcation is local at $\Omega^{-1} = 0$.}
\end{figure}

The low amplitude basin does not exhibit nonlocal geometry upon
approach to the bifurcation at zero $\Omega^{-1}$, and
correspondingly, the linear scaling is not found in down
$\rightarrow$ up transitions at any $\Omega^{-1}$ (see
Fig.~\ref{fig:down-up crossover}), which is in accord with the DSS
theorem.  The $\xi=3/2$ scaling gives way to a different scaling
relationship further from the bifurcation, with an exponent $\xi
\approx 1.3$. 
We interpret this $\xi \approx 1.3$ exponent as a power-law
interpolation.

Note that region III in Fig.~\ref{fig:Phase diagram down-up} (the
$\xi=3/2$ region) extends all the way to zero $\Omega^{-1}$ - this
is unlike section II, where the $\xi=3/2$ region disappears as
$\Omega^{-1} \rightarrow 0$. These results are in accord with the
theorem of DSS which states that close to a bifurcation which is
local at $\Omega^{-1}=0$, in the underdamped regime at non-zero
$\Omega^{-1}$ the $\xi$ is expected to be $3/2$.  Since the
underdamped portion of the hysteresis extends all the way to the
bifurcation as $\Omega^{-1} \rightarrow 0$, we expect the $3/2$
scaling to extend all the way to the bifurcation in this limit.
Had we assumed that $\xi = 3/2$ due to the over-damped soft mode,
we would follow the nearly-adiabatic analysis of the previous
section and we would find that for low $\Omega^{-1}$ there exists
another dimensionless parameter $\nu = \tilde{\eta}^{1/2}\Omega^2$
\footnote{Unlike the previous section, parameters $\tilde{\eta}$
and $\Omega^{-1}$ collapse to such single parameter $\nu$ only up
to second order in $\nu$.  At higher orders, terms are multiplied
by an additional $\Omega^{-2}$.}. When $\nu \ll 1$ such analysis
predicts that $\xi = 3/2$. 
When $\nu \sim 1$, the reduction to 1-d due to adiabaticity
becomes impossible and prediction based on this method can not be
made. The fact that $\xi$ remains $3/2$ beyond $\nu \ll 1$, as we
see from numerics, indicates that the $3/2$ scaling is not a
consequence of reduction to an over-damped 1-d soft-mode.  

In the previous section we offered three definitions of the $3/2
\rightarrow 1$ crossover.  In the case of up $\rightarrow$ down
transitions, we don't have theory for the $\xi \approx 1.3$
scaling regime, so the crossover from the $3/2$ scaling can be
defined here only as a point at which the $3/2$ scaling begins to
break.  Since the $3/2$ scaling is predicted by the
quasi-Hamiltonian theory, we look at next order corrections due to
Chinarov et. al \cite{Chinarov}.  Translating the results of that
work to our notation, close to the bifurcation, $R \approx
\frac{4}{3^{1/4}}\Omega^{-1}\tilde{\eta}^{3/2} + 1.06
\frac{4}{3^{1/4}}\Omega^{-3}\left|\ln{\Omega^{-1}}\right|\tilde{\eta}^{3/2}$
(compare with Eq.~\ref{eq:quasi-ham theory assymp}). Evidently,
the correction is to simply shift $R(\tilde{\eta})$ by a constant
offset (see Fig.~\ref{fig:down-up crossover} (b) where this effect
is clearly pronounced throughout the entire $R(\tilde{\eta})$, not
just close to the bifurcation). Because the correction is so small
for small $\Omega^{-1}$, its effect doesn't become noticeable
until a rather large $\Omega^{-1}$ - equating the zeroth-order
term to the correction term predicts this value of $\Omega^{-1}$
to be $\approx 1.5$. Qualitatively, this delayed effect is
observed in numerics (see Fig.~\ref{fig:Phase diagram down-up}),
but the crossover doesn't seem to depend on $\Omega^{-1}$
significantly until $\Omega^{-1} \approx \Omega^{-1}_c/2  \approx
0.28$.
\section{Conclusion}
We considered the escape rate $W \propto \exp{\left(-R/D\right)}$
from one basin of a non-linear oscillator into another.  We
calculated the leading term in the effective barrier $R$ in the
limit of small noise: $\ln{W} =  -R(\Omega^{-1},\sqrt{\beta})/D +
O(D^0)$. For any given $\Omega^{-1}$, this $R$ may exhibit several
scaling behaviors versus $\left|\sqrt{\beta}-
\sqrt{\beta_B}\right|$ where $\sqrt{\beta_B}$ is a value of the
driving field at which a saddle-node bifurcation takes place. We
used numerical shooting method to calculate most probable escape
paths (MPEPs) from one basin to another. $R$ is the action along
this MPEP between an attractor inside one of the basins and the
saddle that lies on the separatrix between the two basins. This
technique allowed us to map out the $\sqrt{\beta}$ -
$\Omega^{-1})$ parameter space for locations of various scaling
behaviors of $R$ versus $\left|\sqrt{\beta}-
\sqrt{\beta_B}\right|$.  The result of this mapping is shown in
Fig.~\ref{fig:Phase diagram up-down} for up $\rightarrow$ down
transitions and in Fig.~\ref{fig:Phase diagram down-up} for down
$\rightarrow$ up transitions.

We confirmed predictions of the recent work by Dykman, Schwartz
and Shairo (DSS) \cite{Mark and Ira}, which classifies scaling
laws based on the structure of the homoclinic trajectory at the
energy of the saddle zero $\Omega^{-1} = 0$ - "local" or
"nonlocal".  We found that for those bifurcations which become
nonlocal when $\Omega^{-1}=0$, at non-zero $\Omega^{-1}0$ there
exists a regime with $R \propto \left|\sqrt{\beta}-
\sqrt{\beta_B}\right|$, as predicted by DSS. Such situation occurs
for up $\rightarrow$ down transitions close to the low-field
bifurcation.  Even closer to the low-field amplitude bifurcation
we found a crossover into the $R \propto \left|\sqrt{\beta}-
\sqrt{\beta_B}\right|^{3/2}$ regime. We have defined this
crossover in three different ways. Following one of the
definitions we discovered that the $3/2$-scaling breaks down much
further from the bifurcation then where it would be expected to
break down if the physics of this scaling was due to an
over-damped soft mode.

We also found that for those bifurcations which are local when
$\Omega^{-1}=0$, at non-zero $\Omega^{-1}$ there exists a regime
with $R \propto \left|\sqrt{\beta}- \sqrt{\beta_B}\right|^{3/2}$,
which holds over a finite potion of a hysteresis even as
$\Omega^{-1} \rightarrow 0$.  Such situations take place for down
$\rightarrow$ up transitions close to the high-field bifurcation
(this has been predicted earlier \cite{Dmittriev and D'yakonov}).
This $3/2$ scaling again takes place much further from the
bifurcation then were it would be expected to hold as a result of
over-damped soft mode,
but for the down $\rightarrow$ up transitions this fact is
somewhat less novel because it has been explicitly predicted to
hold in the limit of $\Omega^{-1} \rightarrow 0$ a long time ago
by \cite{Dmittriev and D'yakonov} and also by recent work of DSS
as explained in Section II.B.
Further from the high-field bifurcation, we found another
crossover into a regime $R \propto \left|\sqrt{\beta}-
\sqrt{\beta_B}\right|^{\xi}$ where $\xi \approx 1.3$. A crossover
into power law regime was not expected on any analytical grounds,
so we interpret this power law as an interpolation.

The author expresses deep gratitude to Mark Dykman for his
guidance on this project.  A short publication with Dr. Dykman
will appear elsewhere.  The author also thanks Michael Cross for
numerous useful discussions and acknowledges the support from the
NSF Grant No. DMR-0314069.

\end{document}